\documentclass[final]{elsart4-1}

\usepackage{graphicx}
\usepackage{amssymb}

\usepackage[usenames,dvipsnames,svgnames,table]{xcolor}
\usepackage[english,francais]{babel}
\usepackage[utf8]{inputenc}
\usepackage{url}
\usepackage{booktabs}
\usepackage[T1]{fontenc}

\usepackage[numbers]{natbib}

\newcommand{\ra}[1]{\renewcommand{\arraystretch}{#1}}
\def\degr{\hbox{$^\circ$}}
\def\ls{LS~5039}
\def\hessj{HESS~J0632+057}

\def\lsi{LS~I~+61\degr303}
\def\psrb{PSR~B1259-63}
\def\cyg{Cyg~X-3}
\def\la{\mathrel{\hbox{\rlap{\hbox{\lower4pt\hbox{$\sim$}}}\hbox{$<$}}}}
\def\ga{\mathrel{\hbox{\rlap{\hbox{\lower4pt\hbox{$\sim$}}}\hbox{$>$}}}}

\begin{document}
% Select a primary header Physics or Astrophysics
% You can place after the header (classification), if you know it.

\centerline{Astrophysics}
\begin{frontmatter}



\selectlanguage{english}
\title{Gamma-ray emission from binaries in context}


\author{Guillaume Dubus}
\ead{Guillaume.Dubus@obs.ujf-grenoble.fr}
\address{Univ. Grenoble Alpes, IPAG, F-38000 Grenoble, France}
\address{CNRS, IPAG, F-38000 Grenoble, France}

%\medskip
%\begin{center}
%{\small Received *****; accepted after revision +++++}
%\end{center}

\begin{abstract}
More than a dozen binary systems are now established as sources of variable, high energy (HE, $0.1-100\,\rm GeV$) gamma rays. Five are also established sources of very high energy (VHE, $\geqslant 100\,$GeV) gamma rays.  The mechanisms behind gamma-ray emission in binaries are very diverse. My current understanding is that they divide up into four types of systems: gamma-ray binaries, powered by pulsar rotation; microquasars, powered by accretion onto a black hole or neutron star; novae, powered by thermonuclear runaway on a white dwarf; colliding wind binaries, powered by stellar winds from massive stars. Some of these types had long been suspected to emit gamma rays (microquasars), others have taken the community by surprise (novae). My purpose here is to provide a brief review of the current status of gamma-ray emission from binaries, in the context of related objects where similar mechanisms are at work (pulsar wind nebulae, active galactic nuclei, supernova remnants).

\vskip 0.5\baselineskip

\selectlanguage{francais}
\noindent{\bf R\'esum\'e}
\vskip 0.5\baselineskip
\noindent 
{\bf Emission gamma des systèmes binaires.}
Plus d'une douzaine de systèmes binaires sont maintenant identifiés comme des sources variables de rayonnement gamma de haute énergie ($0.1-100\,\rm GeV$). Cinq systèmes binaires sont également des sources de rayonnement gamma de très haute énergie ($\geqslant 100\,$GeV). Les processus menant à l'émission de ce rayonnement sont variés. On peut néanmoins diviser ces systèmes  en quatre grandes classes: les binaires gamma, dont le moteur est la rotation d'un pulsar; les microquasars, dont le moteur est l'accrétion sur un trou noir ou une étoile à neutrons; les novae, dont l'énergie provient de la combustion nucléaire à la surface d'une naine blanche; les binaires à collision de vents, qui dissipent l'énergie cinétique de vents d'étoiles massives. On soup\c connait depuis longtemps que certaines classes de systèmes binaires devaient être associées à de l'émission gamma (les microquasars); pour d'autres classes, la découverte d'émission gamma a été une surprise (les novae). Je propose ici une brève revue de nos connaissances sur l'émission gamma de haute et très haute énergie dans les systèmes binaires, avec le souci de souligner les liens avec d'autres objets astrophysiques dans lesquels des processus similaires sont à l'oeuvre (nébuleuses de pulsar, noyaux actifs de galaxie, restes de supernova).
%Now keywords/mots-cls
%\keyword
\vskip 0.5\baselineskip
\noindent{\small{\it Keywords~:} Acceleration of particles;  Radiation mechanisms: non-thermal; Binaries: general; Gamma-rays: stars}
\vskip 0.5\baselineskip
\noindent{\small{\it Mots-cl\'es~:} Accélération de particules; mécanismes de rayonnement non-thermiques; étoiles binaires; rayonnement gamma: étoiles}

\end{abstract}

%\begin{keyword}
     
%\end{keyword}

\end{frontmatter}

% \linenumbers

% now the Version française abrégée, if it exists
%\selectlanguage{francais}
%\section*{Version fran\c{c}aise abr\'eg\'ee}
% Text of your Version française abrégée here

\selectlanguage{english}

\section{Introduction}
\label{intro}
Detections of binaries punctuated high energy and very high energy gamma-ray astronomy in the 1970s and 1980s. Binaries had been discovered to be the brightest sources in the newly-opened X-ray sky and were thus likely candidates for emission at even higher energies. Most of these detections did not withstand the test of time as independent confirmation could not be obtained \cite{Weekes:1992lf}. Yet, they fostered great interest and participated in the development of the field from the 1980s onwards. 

The latest generation of gamma-ray instruments, from the mid-2000 onwards \citep{cras-fermi,cras-ground},  have brought back binaries to the forefront. Many binaries are now established sources of high energy (HE, $0.1-100\,\rm GeV$) and very high energy (VHE, $\geqslant 100\,$GeV) gamma rays: I have listed in Table 1 the names and main characteristics of those that are reported as secure gamma-ray detections {\em and} display variability on a timescale that can be related to the binary nature of the source. I have also listed one or two references to the most recent HE and/or VHE studies as a point of entry into the litterature.  Some of these binary gamma-ray sources are of the type that had long been suspected to emit gamma rays (microquasars), others have taken the community by surprise (novae).  There is no simple extrapolation between the HE and VHE domains. At present, the only binaries detected in VHE gamma rays appear to be composed of a pulsar and a massive star. There is much more diversity in HE gamma rays, perhaps because the higher fluxes and the instrumental sensitivity at these energies give access to more objects.

My purpose here is to sketch a status of gamma-ray emission from binaries as of early 2015 and how it is understood in the broader context of related objects and high-energy emission models. More in-depth accounts of the observations and models may be found in other reviews \cite{Bosch-Ramon:2008hg,2013APh....43...81B,2013A&ARv..21...64D,2014arXiv1410.3758L}. My current understanding is that the binaries detected in HE or VHE gamma rays divide up into four classes, illustrated in Figure 1:
\begin{itemize}
\item {\it gamma-ray binaries}, powered by pulsar rotation;
\item {\it microquasars}, powered by accretion onto a black hole or neutron star;
\item {\it novae}, powered by thermonuclear runaway on a white dwarf;
\item {\it colliding wind binaries}, powered by stellar winds from massive stars.
\end{itemize}
The last three classes were well-established prior to the detection of gamma rays. Gamma-ray binaries are still rather new and, although some of the systems had been known for some time, the motivation to treat them as a separate class is an outcome of the last decade of gamma-ray observations. The contours of this class are not fixed: sometimes the term ``gamma-ray binary'' is used in the literature to designate all binaries that have been detected in HE or VHE gamma rays (all the above), sometimes it encompasses only the systems listed as high-mass gamma-ray binaries in Table 1, sometimes it adds the microquasars to the high-mass gamma-ray binaries. I think the first definition is too broad, the second turns out to be too restrictive, the third mixes objects with very different phenomenology and underlying powering mechanisms.

A distinguishing feature of a {\em gamma-ray} binary is that emission above 1 MeV dominates its spectral energy distribution (see example in Figure 2), disregarding the black body like component from the companion star \cite{2013A&ARv..21...64D}. Another distinguishing feature of the gamma-ray binaries listed in Table 1 is that they all have variable gamma-ray emission, sometimes modulated on the orbital period. This definition encompasses both binaries in which the companion to the compact object is a high-mass  ($\geqslant 10\,\rm M_\odot$) star and those where the companion is low-mass ($\leqslant 1\,\rm M_\odot$). The ensuing subdivision into low-mass and high-mass gamma-ray binaries, like X-ray binaries, actually makes sense as I shall explain in section 2.  Radio pulsations identify the compact object as a neutron star in many of these systems.  Theoretically, the prevailing idea is that gamma-ray emission from these systems is powered by the rotation of the pulsar magnetosphere. 

X-ray binaries are powered by accretion of matter from a normal star onto a stellar-mass black hole or a neutron star. Infalling matter forms an accretion disk around the compact object, where its gravitational energy is gradually released. In some X-ray binaries, part of this energy ends up powering a  jet {\em i.e.} a collimated ejecta with relativistic speeds, similar to those seen in active galactic nuclei (radio quasars, blazars), hence the term ``microquasar'' \cite{Fender:2002hk}.  Most of the radiated energy is released in X-rays. The prevailing idea is that the gamma-ray emission is associated with the formation of the relativistic jet. 

Novae events are due to runaway thermonuclear burning of material deposited on the surface of a white dwarf.  Novae occur in cataclysmic variables, binaries with a white dwarf accreting from a stellar companion  \cite{Warner:1995mo}. Depending upon the mass transfer rate, it can take tens to thousands of years for the accumulated material to reach the critical pressure that triggers thermonuclear burning. The prevailing idea is that the gamma-ray emission is associated with the ejection of part of the material when runaway burning starts. The rest of the material inflates into a large enveloppe around the white dwarf until, after days to weeks, nuclear burning ceases and the pressure is insufficient to maintain the enveloppe. Most of the radiated energy is released in the optical band where they are sometimes visible to the naked eye. 

Colliding wind binaries are composed of massive O or Wolf-Rayet stars. These stars have masses in excess of 20$\,\rm M_\odot$ and large luminosities $10^4-10^5\,\rm L_\odot$. The radiation pressure drives strong winds from the stars with terminal velocities $\approx 1000-2000\rm\,km\,s^{-1}$ and mass loss rates of $10^{-8}-10^{-3}\,\rm M_\odot\,yr^{-1}$. The collision between the stellar winds in a binary forms a shock structure where the released kinetic energy heats the gas to $\sim 10^7$~K, which emits principally X-rays \cite{2005xrrc.procE2.01P,2007A&ARv..14..171D}. Some of the energy may be channeled into accelerating particles at the shock, as in supernovae remnants, leading to gamma-ray emission.

The detection of gamma rays has brought new light, literally, to the mechanisms in operation in these systems. The mechanisms behind gamma-ray emission in binaries are very diverse (see Figure 1 for illustrations of these). In novae, as will be explained in section 4.2, the physics has close analogies to that of supernova remnants;  the gamma-ray emission from microquasars involves relativistic jets as active galactic nuclei and gamma-ray bursts do; many binaries host pulsars whose rotating magnetic field ends up powering strong non-thermal emission. Table 1 makes it clear that the binaries emitting gamma rays do not constitute a class of objects united by a common astrophysical model, as for instance supernova remnants, but by a common phenomenon: orbital motion. Orbital motion constraints the size and type of the binary components, and allows measurements to probe different conditions or vantage points. The legitimacy of theories developed for gamma-ray sources in other astrophysical contexts can thus be put to the test in the peculiar environment of binaries.

In the following, I have tried to emphasize how  gamma-ray observations of binaries provide such tests. Thus, section 2 sets gamma-ray binaries in the context of pulsars, section 3 relates microquasars to the physics of relativistic jets, and section 4 examines the gamma-ray detections of novae and colliding winds binaries in the context of particle accretion at non-relativistic shocks. 

\begin{table}
%\begin{sidewaystable}
\centering
{\small%\footnotesize
\ra{1.27}
\begin{tabular}{@{}lcclccll@{}}
\toprule 
name & \multicolumn{2}{c}{~~~~~binary components~~~~~}  & P$_{\rm orb}$ (d)~  & ~HE~ & ~VHE~ & ~refs ($\star$)~ & ~notes  \\
%	 &  &  &  (day) & \multicolumn{2}{c}{$\gamma$-ray}   &  \\
 \toprule 
\multicolumn{8}{l}{\cellcolor{Lavender}(high-mass) gamma-ray binaries}\\
 %\multicolumn{7}{l}{\cellcolor{Lavender}{gamma-ray binaries}}\\
 \midrule
PSR B1259-63 	 	& pulsar 	& Be & 1236.7 	& \checkmark & \checkmark & \cite{2011ApJ...736L..11A,H.E.S.S.Collaboration:2013fk} & 47.7 ms \\
HESS J0632+057	& ? 		& Be & 315  		&  & \checkmark & \cite{2013MNRAS.436..740CBIS,2014ApJ...780..168A} & \\
LS I +61$^{\circ}$303& ? 		& Be & 26.5 		& \checkmark & \checkmark & \cite{2011ApJ...738....3A,2013ApJ...773L..35A} & magnetar ? \\
1FGL J1018.6-5856	& ? 		& O & 16.6 	&  \checkmark & \checkmark & \cite{2012Sci...335..189F,2015AA...577A.131H} &\\
LS 5039 		 	& ? 		& O & 3.9		& \checkmark & \checkmark & \cite{Aharonian:2006qw,2012ApJ...749...54HBIS} &\\
 %$t_{0}$ 			& 48124.34911(9) MJD	& 53478.09(6) JD 	& 54857.5 HJD (per 0.967$\pm$0.008)  & 55403.3(4) MJD\\
 %$e$ 				& 0.8698872(9) 	&  0.35(3) 		& 0.54(3) 		& 0.83(8) 		& - \\
 %$\omega$ (\degr)	& 138.6659(1)$^\sharp$ 		&  212(5) 		& 41(6) 		& 129(17)  & - \\
 %$i$ (\degr)		& 19--31 			& 13--64 			& 10--60 		& 47--80 		& -\\
 %$d$ (kpc)			& 2.3(4) 			& 2.9(8) 			& 2.0(2) 		& 1.6(2)  		& 5.4\\
\midrule 
 \multicolumn{8}{l}{\cellcolor{Lavender}{(low-mass) gamma-ray binaries (${\dagger}$)}}\\
 \midrule
XSS J12270-4859 	& pulsar 	& red dwarf & 0.29	& \checkmark & & \cite{2015ApJ...800L..12RBIS,2015ApJ...806...91JBIS} &1.7 ms\\
PSR J1023+0038 	& pulsar 	& red dwarf & 0.20 	& \checkmark & & \cite{2014ApJ...790...39S} & 1.7 ms\\
2FGL J0523.3-2530 & ? & red dwarf & 0.69 & \checkmark &  & \cite{2014ApJ...788L..27S,2014ApJ...795...88X} &\\
PSR B1957+20 	& pulsar   & brown dwarf & 0.38 & \checkmark & & \cite{2012ApJ...761..181W} & 1.6 ms \\
PSR J0610-2100	& pulsar	& brown dwarf & 0.29 & \checkmark & & \cite{2013MNRAS.430..571E} & 3.8 ms \\
PSR J1311-3430	& pulsar 	& brown dwarf & 0.065 & \checkmark & & \citep{2012ApJ...754L..25R,2015ApJ...804L..33X} & 2.6 ms \\
%1FGL J1417.7-4407 & pulsar ? & red giant & 5.4 & \check mark & & \citep{2015ApJ...804L..12S} \\
%PSR J1824?2452I	& pulsar 	& red dwarf & 0.45 	& \checkmark & &  (M28I) 3.9 ms\\
\midrule 
 \multicolumn{8}{l}{\cellcolor{Lavender}microquasars (X-ray binaries)}\\
 \midrule
Cyg X-3 	 	& black hole ? 	& Wolf-Rayet & 0.20 	& \checkmark & & \cite{2012MNRAS.421.2947CBIS,2010ApJ...721..843A} &\\
Cyg X-1 	 	& black hole~~ 	& O & 5.60 	& \checkmark & ? & \cite{2013arXiv1305.5920M,Albert:2007uw} &\\
\midrule 
 \multicolumn{8}{l}{\cellcolor{Lavender}novae}\\
 \midrule
V407 Cyg 	 	& white dwarf 	& red giant & 14000 ?& \checkmark & &\cite{2010Sci...329..817A,2012ApJ...754...77A}  &N Cyg 2010\\
V1324 Sco	 	& white dwarf 	& red dwarf & 0.07 ? 	& \checkmark & & \cite{2014Sci...345..554A} & N Sco 2012\\
V959 Mon	 	& white dwarf 	& red dwarf & 0.30 & \checkmark & & \cite{2014Sci...345..554A} & N Mon 2012 \\
V339 Del 			& white dwarf 	& red dwarf & 0.13 ? 	& \checkmark & &\cite{2014Sci...345..554A} & N Del 2013 \\
V1369 Cen 		& white dwarf 	& red dwarf & ? & \checkmark & & \cite{2013ATel.5653....1C} & N Cen 2013 \\
%V745 Sco 	& white dwarf 	& M giant & * d 	& \cellcolor{LightSteelBlue}\checkmark & & Nova Sco 2014  Banerjee 2014\\
\midrule 
 \multicolumn{8}{l}{\cellcolor{Lavender}colliding wind binary}\\
 \midrule
Eta Car	 	& LBV 	& O/WR ? & 2014 	& \checkmark & & \cite{2012AA...544A..98R,2012MNRAS.424..128H} & \\
%\bottomrule 
%\\
\multicolumn{8}{l}{$\star$ I only give one or two recent references as entry points to the HE/VHE litterature.}\\
\multicolumn{8}{l}{$\dagger$ not including another $>50$ {\em Fermi}-LAT pulsars in binaries.}\\
\end{tabular}
}
\caption{Binary systems identified with sources of variable HE or VHE gamma-ray emission\label{binaries}.}
\end{table}

\begin{figure}
\centering\resizebox{0.6\hsize}{!}{\includegraphics{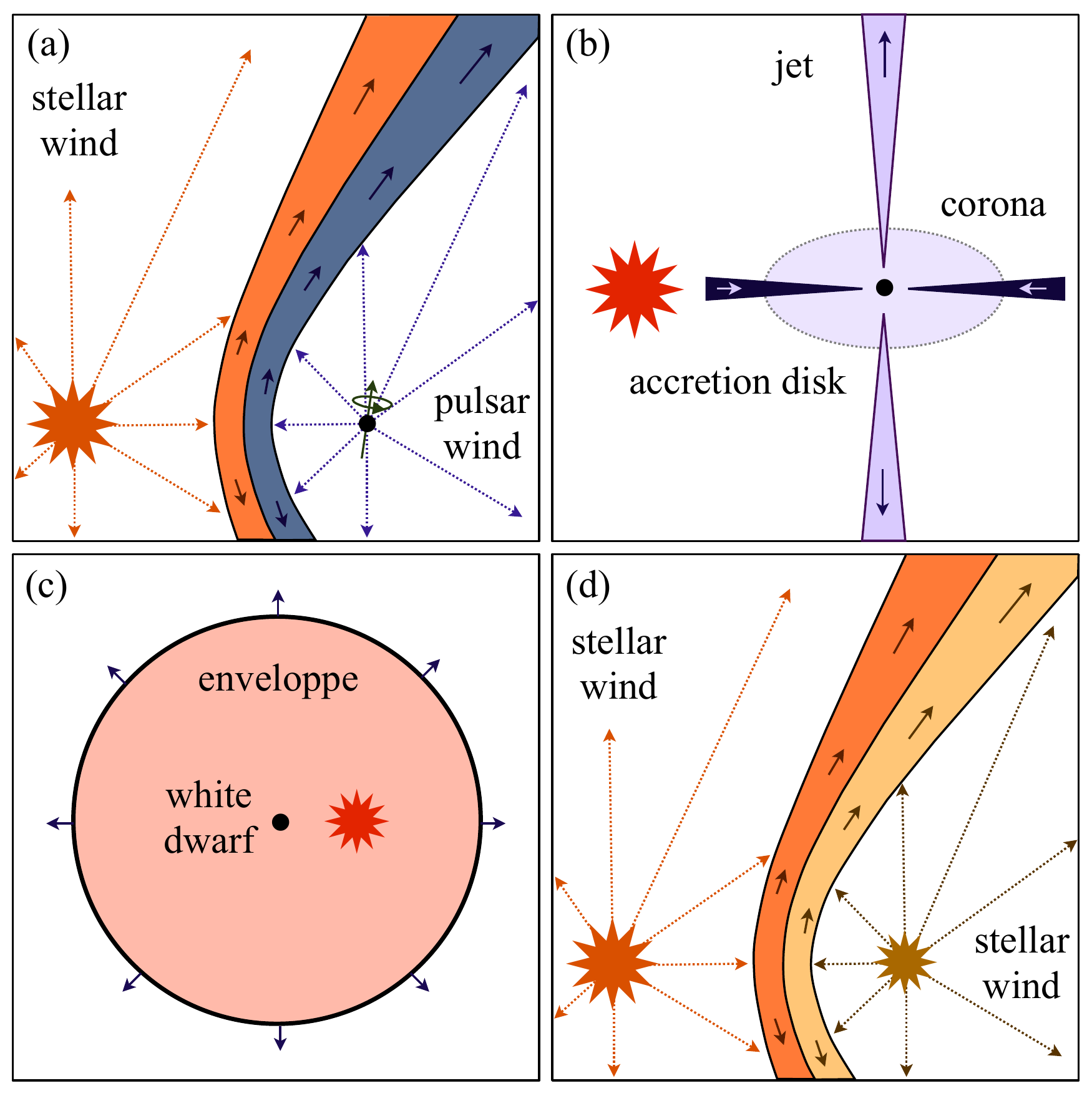}}
\caption{Sketch of the types of gamma-ray emitting binaries. (a) A high-mass gamma-ray binary, with a rotation-powered pulsar. The filled in region represents the shocked pulsar wind and the shocked stellar wind (panel adapted from \cite{2013A&A...557A.127D}). (b) A microquasar, with its accretion disk + corona, and the relativistic jet. (c) A nova, with the expanding ejecta and enveloppe around the white dwarf. (d) A colliding wind binary, with two stellar winds interacting. The sketches are not to scale.}
\end{figure}

\section{Gamma-ray binaries and pulsars\label{bp}}

\subsection{\psrb\ as a binary PWN}
The detection of \psrb\ with the H.E.S.S. arrays of Cherenkov telescopes was the first detection of a binary with the latest generation of gamma-ray instrumentation \cite{2004ATel..249....1B}. As the name implies, \psrb\ is a binary system harbouring a pulsar. Its 48.8\,ms pulsations were detected with the Parkes radio telescope whilst surveying the sky for new pulsars \cite{1994MNRAS.268..430J}. The radio pulsations can be tracked very accurately, allowing for a precise determination of the orbit and the measurement of the rate at which the pulsar rotation slows down. The slow down implies rotational kinetic energy is lost at a rate $\dot{E}\approx 8\times 10^{35}\rm\,erg\,s^{-1}$ ($8\times 10^{28}\rm\,W$), which is a high spindown power for a pulsar.  This energy loss basically corresponds to the power required to maintain the magnetic field in rotation with the neutron star. At zeroth order the pulsar behaves like a rotating dipole, enabling the pulsar spindown power to be related to the magnetic field. In \psrb\, the magnetic field of the pulsar is $B\approx 3.3\times10^{11}\rm\,G$ ($3.3\times10^{7}\rm\,T$). These characteristics make \psrb\ a rather typical {\em rotation-powered} young pulsar \cite{2012puas.book.....L}.

The VHE and HE gamma-ray emission are not pulsed but are only detected close to periastron passage, when the pulsar is closest to its companion. The companion is a massive star with a mass $\approx 30\rm\,M_\odot$ \cite{2011ApJ...732L..11N}.  In addition to the fast, radiatively-driven stellar wind, spectroscopy shows emission lines arising from a slowly outflowing disk of material in Keplerian rotation around the equator of the star. The origin of these circumstellar disks, characteristic of ``Be stars'', is not known but  thought to be due to the star rotating close to breakup velocity \cite{2013arXiv1310.3962R}. The 3.4 year orbit has an eccentricity of 0.87 that takes the pulsar from 13 AU to 0.9 AU from the star. At periastron, the pulsar is about 20 stellar radii away from the star. The density of material in the circumstellar disk and stellar wind is then high enough to eclipse the radio pulsations for $\approx 40$\,days around periastron due to free-free absorption \cite{1995MNRAS.275..381M}. 

In rotation-powered pulsars, most of the energy lost by pulsar spindown powers a tenuous, highly relativistic wind carrying $e^+ e^-$ pairs and electromagnetic energy away beyond the pulsar light cylinder radius $R_L$ \cite{Kirk:2007wh,Arons:2011aa}, the radius at which the rotation velocity of material tied to the pulsar magnetic field lines reaches the speed of light and must then decouple ($R_L=c P/2\pi$ where $P$ is the pulsar spin period). When the pulsar is isolated, the pulsar wind propagates freely until it is contained by the interstellar medium or the ejecta from its parent supernova,  forming a shock structure. These are called pulsar wind nebulae (PWN) \cite{Gaensler:2006qi,cras-pwn}. In \psrb, the pulsar wind propagates in the much denser environment provided by the circumstellar disk and stellar wind. As a result, the pulsar wind is (at least partly) confined, forming a shock structure with the surrounding stellar material on a much smaller scale than in a PWN \cite{Tavani:1994qu}.

The distance to which the pulsar wind propagates freely is estimated by equating the pulsar wind ram pressure to the pressure of the surrounding material. If the surrounding (ram) pressure is due to an isotropic, coasting stellar wind, the distance $R$ to the pulsar is 
\begin{equation}
\frac{R}{d}\approx \frac{1}{1+\eta^{1/2}}~~{\rm with}~~\eta=\frac{\dot{M}_{\rm w} v_{\rm w}}{\dot{E}/c}
\end{equation}
where $d$ is the orbital separation, $\dot{E}$ is the pulsar wind power, $\dot{M}_{\rm w}$ and $v_{\rm w}$ the stellar wind mass loss rate and velocity (see \cite{2013A&ARv..21...64D} for details and references). A stellar wind with a strong `thrust' $\dot{M}_{\rm w} v_{\rm w}$ contains the pulsar wind close to the neutron star ($\eta\gg1 \Rightarrow R/d\ll 1$). Inversely, a strong pulsar wind pushes back the stellar wind all the way to the surface of the star ($\eta\ll 1 \Rightarrow R/d\approx 1$). 

Here, as in PWN \citep{cras-pwn}, the shock is thought to channel a fraction of the pulsar wind energy to particles that then radiate gamma rays. The process may involve randomization of the particles initially frozen with the magnetic field in the pulsar wind, diffuse shock acceleration, or shock-driven reconnection. The latter is increasingly scrutinized since it has become clear that highly relativistic pair plasmas are poor at shock-accelerating electrons to very high energies unless they are very weakly magnetized (probably too much so to be reconciled with observations and with the requirement that the pulsar wind starts at the light cylinder by carrying a lot of magnetic energy) \citep{cras-acc}. Anyhow, we know that the process must be efficient because a large fraction of the spindown power ends up as non-thermal radiation: the peak luminosity in HE gamma rays from \psrb\ is nearly equal to its spindown power \cite{2011ApJ...736L..11A} !

Besides \psrb, there are four other known pulsars with massive star companions: PSR J1740-3052, PSR J1638-4725, PSR J0045-7319, and PSR J2032+4127. The first three are probably too faint in gamma rays to be detectable because of their lower spindown power and larger distance than \psrb\  \citep{2013A&ARv..21...64D}. The last one, PSR J2032+4127, has 143\,ms pulsations detected in radio and HE gamma rays and has been tentatively associated with VHE source TeV J2032+4130  \citep{2014ApJ...783...16A}. PSR J2032+4127 is actually in a long  (20--30 year), highly eccentric ($e\approx 0.95$) orbit with a 15\,$M_\odot$ star \citep{2015MNRAS.451..581L}. Like \psrb, variable gamma-ray emission may be detectable in early 2018 when periastron passage  will bring the pulsar within 2 AU ($3\times 10^{11}\,m$) of its companion star.

Four other binaries, listed in Table 1, currently have observational properties that are similar to \psrb\ (modulated HE or VHE gamma-ray emission, massive star companion, etc). Based on this, the binary PWN scenario described above for \psrb\ has been widely adopted for those four systems, even if some contest this interpretation, favoring a microquasar scenario \cite{2012A&A...540A.142M}. The basic issue is that it has not been possible to detect radio pulsations, most likely because these binaries have much shorter orbital periods than \psrb. The pulsar is so deeply embedded within the dense stellar wind that the radio signal cannot propagate out \cite{Dubus:2006lc}. Still, two magnetar bursts in X-rays have been detected from the direction of \lsi, within $\approx $ 2 arc minutes of the source, strengthening the case for a pulsar in this system \cite{2012ApJ...744..106T}.  I am of the opinion that the array of evidence strongly favours the binary PWN scenario and refer to \cite{2013A&ARv..21...64D} for a detailed discussion.

\subsection{Low-mass gamma-ray binaries}
Gamma-ray emitting pulsars are also found in association with low-mass companions: red dwarfs, brown dwarfs, white dwarfs, other pulsars. For instance, HE gamma rays were detected from the famous double pulsar PSR J0737-3039A/B \cite{2013ApJ...768..169GBIS}. As of November 2014, 55 of the 161 pulsars in the {\em Fermi-LAT} catalog of gamma-ray pulsators\footnote{\url{https://confluence.slac.stanford.edu/display/GLAMCOG/Public+List+of+LAT-Detected+Gamma-Ray+Pulsars}} are actually in binaries, far outnumbering the systems listed in Table 1. These pulsars have millisecond pulse periods and belong to the class of old pulsars that have been ``recycled" {\it i.e.} spun back up by accretion of material from the companion onto the neutron star \cite{2012puas.book.....L}. In nearly all of these systems, the only difference with an isolated gamma-ray pulsar is that the pulse frequency is modulated by the orbital motion of the pulsar \citep{cras-pulsars}. 

However, there are a few of those systems where their presence in a binary has an impact on the gamma-ray emission. These are the five systems that I have listed in Table 1 as low-mass gamma-ray binaries, because gamma-ray emission dominates their spectral energy distribution, they show variability linked to their binary nature (albeit often tentatively), and they also involve interacting pulsars.

The HE gamma-ray emission from PSR B1957+20, PSR J0610-2100, PSR J1311-3430, and 2FGL J0523.3-2530 shows some evidence for a modulation on the orbital period \cite{2012ApJ...761..181W,2013MNRAS.430..571E,2014ApJ...795...88X}. PSR B1957+20 is the original ``black widow'' pulsar, named as such because the pulsar wind in this system impinges directly onto its very low mass companion ($\leqslant 0.1\rm\,M_\odot$), ablating material from the star (see \cite{2014JASS...31..101H} for a review). There is no strong enough stellar wind or accretion flow present to stop the pulsar wind ($\eta\ll 1$).  Part of the gamma-ray spectrum is not variable and attributed to pulsar magnetospheric emission, as in isolated pulsars, while the emission above $\approx 2\,\rm GeV$ is variable and attributed to the pulsar wind. The processes are identical to those invoked for the ``high-mass'' gamma-ray binaries ({\em e.g.} \cite{Arons:1993di,2014A&A...561A.116B}). The modulation is detected in only part of the {\em Fermi}-LAT data on 2FGL J0523.3-2530, raising the interesting possibility that orbital modulations may also be sporadically present in some of the other 50 or so pulsars in binaries  mentioned above.

PSR J1023+0038 and XSS J12270-4859 are nearly twin systems (companion, orbital period, pulse period), part of the ``redback'' class of pulsars (basically black widows with companion masses of a few tenths instead of a few hundredths of solar mass, see \cite{2014JASS...31..101H}). Both are intermittent HE gamma-ray sources and both show intermittent signs of accretion, manifest as an increase in X-ray and optical/UV flux, stronger variability, and double-peaked optical emission lines typical of accretion disks. Both are sometimes detected as radio pulsars, and sometimes detected as X-ray pulsars \cite{2014arXiv1412.1306A,2014arXiv1412.4252P}. The neutron star seems to transition back and forth between accretion-powered and rotation-powered regime, even combining both at times ! Another such ``transitional millisecond pulsar'' is PSR J1824-2452I in the globular cluster M28 \cite{2013Natur.501..517P}. Although gamma-ray emission is detected from M28, most of the flux seems to arise from another pulsar in the globular cluster \cite{2013ApJ...778..106J} so I have not included it in Table 1. The {\em Fermi}-LAT sources 3FGL J1544.6-1125 and 1FGL J1417.7-4407 also very likely belong to this sub-class of sources \citep{2015ApJ...804L..12S,2015ApJ...803L..27B}. I have not included them in Table 1 because gamma-ray variability (pulsations or flux changes) has yet to secure the identifications.

These transitions are a key verification that old, millisecond radio pulsars have been spun up by accretion. They also represent a challenge since the models predicted that the pulsar wind, once the radio pulsar turned on, would quench accretion and forbid a return to an accreting state \cite{Illarionov:1975yk}. Curiously, in PSR J1023+0038, the HE gamma-ray flux increased by a factor 5 when the source entered an active X-ray state, whereas it was expected that the gamma-ray emission associated with the radio pulsar would disappear with the onset of accretion. A  possibility is that the pulsar mechanism is not turned off and that the additional light from the active accretion disk, which does not extend to the neutron star, leads to higher levels inverse Compton gamma-ray emission from high-energy particles in the pulsar wind \cite{2014ApJ...785..131T}.

\subsection{Gamma-ray binaries and the physics of pulsar winds}

Binaries give access to pulsar winds on smaller scales than in PWN since the termination shock distance is a fraction of an AU (the orbital separation), compared to a fraction of a parsec in a PWN like the Crab nebula. In units of the light cylinder radius $R_{\rm LC}$ (the natural scale for the pulsar wind), both low and high-mass gamma-ray binaries probe scales of $\sim 10^{4} R_{\rm LC}$ instead of $\sim 10^{9} R_{\rm LC}$ for an isolated pulsar \cite{2014AN....335..313R}. The star provides a source of photons that can scatter off the high energy electrons in the pulsar wind and the shocked region. The stellar radiation field is very well known so the resulting up-scattered gamma-ray spectrum constrains the location and properties of the high energy electrons. However, this is  complicated by the anisotropy of the radiation processes involved (inverse Compton emission, pair production) and by relativistic Doppler boosting (the shocked wind maintains mildly relativistic speeds) which, together with the varying orbital separation, contribute to modulate the non-thermal emission up to the gamma-ray regime \cite{2013A&ARv..21...64D}. 

Uncertainties in the geometry of the shock region also  complicate the interpretation. Estimating the exact location of the termination shock is sometimes difficult because the interaction produces a complex, 3D, time-dependent morphology, notably when a circumstellar disk is present. This has begun to be addressed through numerical simulations \cite{2012MNRAS.419.3426B,2012ApJ...750...70T,2012A&A...544A..59B,2013A&A...560A..79L,2015A&A...574A..77P}. Studying systems where the massive star is an O star, without a circumstellar disk, simplifies the geometry and may be more amenable to interpretation. Yet, the observations pose an increasing number of challenges to models. 

First, the gamma-ray spectra show unexpected complexity with two, if not three, distinct emission components in high-mass gamma-ray binaries. For example, in \ls\ (see Figure 2), the MeV spectrum ({\em CGRO}-COMPTEL) does not connect to the GeV spectrum ({\em Fermi}-LAT), itself clearly distinct from the TeV spectrum (H.E.S.S.) \cite{2014A&A...565A..38C}. The gamma-ray fluxes are modulated on the orbital period, but with different phasing (GeV {\it vs} TeV and MeV, see Figure 2). Cross-calibration issues between instruments are not suspected, so this signals different populations of high-energy particles and/or emission mechanisms.Figure 2 shows a simple one-zone model where electrons with a $E^{-2}$ distribution in energy radiate synchrotron and upscatter blackbody photons from the star (at 0.2 AU), illustrating why the GeV emission requires an additional population of electrons\footnote{The magnetic field intensity ($B\approx 0.04\,$T) is set by the ratio of synchrotron to Compton fluxes \citep{cras-acc}. The gamma-ray spectrum fails to reproduce the hard H.E.S.S. spectrum, mostly because it is calculated in the isotropic approximation whereas it should take into account that the incoming photons from the star come from a specific direction as well as relativistic effets \citep{2015arXiv150501026D}. The slope of the synchrotron emission fits well with the X-ray to MeV spectrum, but note that this is not self-consistent since very strong cooling is expected at the high-energy end.}. Various possibilities have been proposed for the origin of the GeV emission. An exciting proposition is that it is the inverse Compton emission of pairs in the pulsar wind \cite{Ball:2000lr,2008A&A...488...37C,2012ApJ...752L..17K,2013A&A...557A.127D}. The pulsar wind is usually `dark' in PWN because the particles are frozen in with the magnetic field and the background radiation field is weak. In binaries, the companion behaves as a torch illuminating the pulsar wind, instilling hope of a direct access to its  energetic and particle content. 

Second, the gamma-ray lightcurves show much more complex behaviour than straightforward `geometrical' radiative models usually account for. These models are based on the change with orbital phase of the binary geometry seen by the observer. Typically, strong gamma-ray emission will be expected when the high-energy electrons are close to the source of seed photons and scatter back the radiation in the direction of the observer ({\em e.g.} \cite{Bednarek:2007qd,Dubus:2007oq}). For instance, in \psrb,  the gamma-ray emission due to inverse Compton scattering should peak shortly before periastron passage if most of the high-energy electrons reside close to the pulsar and most of the seed photons come from the companion star. However, the HE lightcurve actually peaks a month after periastron passage, when the scattering geometry is unfavourable \cite{2013A&A...557A.127D}. The HE luminosity also implies a very high efficiency in converting the spindown power into gamma rays that turns out to be hard to achieve, unless some relativistic boosting is involved \cite{2011ApJ...736L..10T}. Another difficulty raised by the gamma-ray lightcurves are the changes in the HE and VHE modulation of \lsi\ on timescales both longer and shorter than the orbital period \cite{2013ApJ...773L..35A,2014ATel.6785....1H}. Modelling will be difficult if a lot of the gamma-ray variability happens to be related to fluctuations in the circumstellar environment of the pulsar (clumps in the stellar wind, cycles in stellar mass-loss rate, spiral waves in the Be disk, etc \cite{2014arXiv1410.3758L,2014A&A...561L...2Z}). Observations on a long time base are required to extract the average behaviour, and thus identify the regular variations due {\em e.g.} to the periodic compression of the pulsar wind (when the orbit is eccentric) from the more irregular ones due to the companion's weather. 

Finally, robust conclusions on pulsar winds require good knowledge of the binary system parameters: orbital period, eccentricity, orientation, but also stellar wind mass loss rate, velocity and, ideally, the pulsar spin period and spindown power \cite{Sierpowska-Bartosik:2008wt}. This represents a major observational effort, and not all of the parameters are easily accessible in all systems. In light of this, increasing the number of known gamma-ray binaries will help identify the more accessible systems, in addition to contributing to our knowledge of this population.

\begin{figure}
\centering\resizebox{0.49\hsize}{!}{\includegraphics{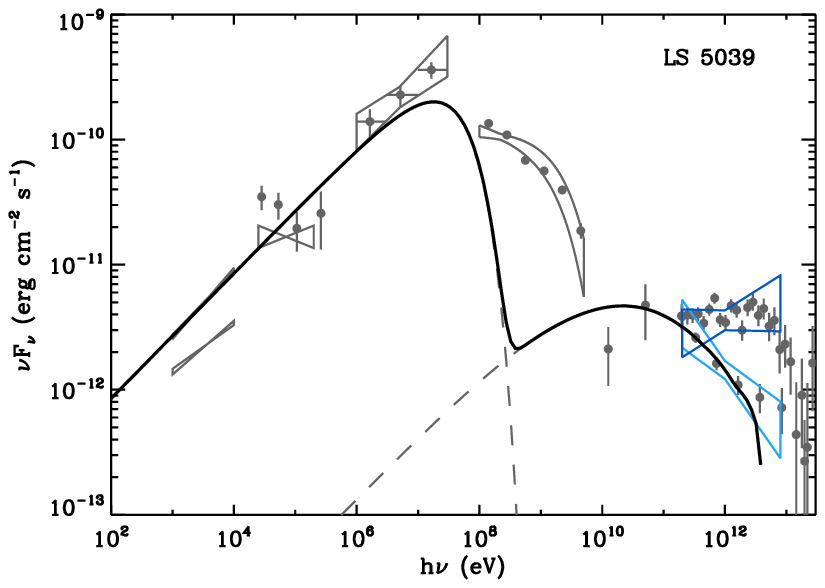}}
\centering\resizebox{0.49\hsize}{!}{\includegraphics{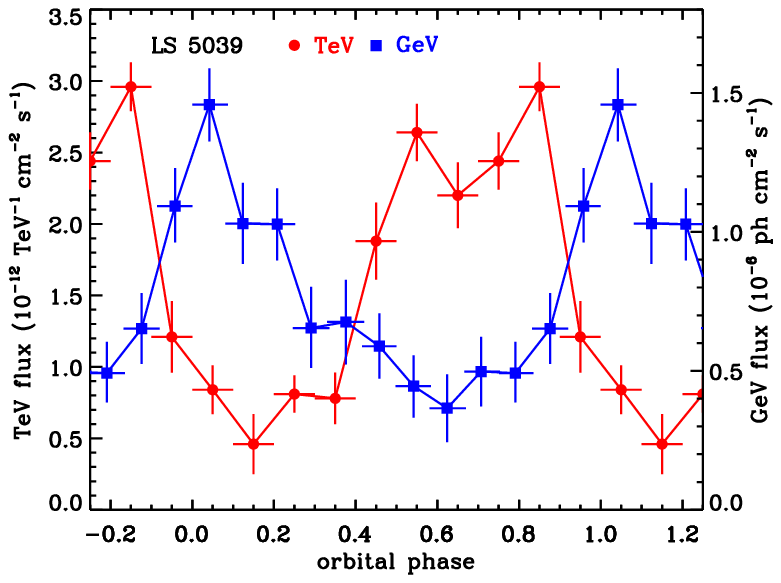}}
\caption{Spectral energy distribution (left) and orbital modulation (right) of the 0.1-10 GeV (blue squares, \citep{2009ApJ...706L..56A}) and TeV (red circles, \citep{Aharonian:2006qw}) flux from the gamma-ray binary LS 5039. The spectrum includes data from Suzaku \citep{Takahashi:2008vuBIS}, BATSE \& INTEGRAL  \citep{Harmon:2004ml,Hoffmann:2008ys}, COMPTEL \citep{2014A&A...565A..38C}, {\em Fermi}-LAT \citep{2012ApJ...749...54HBIS}, H.E.S.S. \citep{Aharonian:2006qw}, in order of increasing frequency. A simple one-zone model is adjusted to the data for illustration purposes, including synchrotron and inverse Compton emission from a single population of electrons (see section 2.3 for details).}
\end{figure}

\section{Microquasars}
\label{bj}
Microquasars, as scaled-down versions of active galactic nuclei (AGNs), have long been on the list of likely HE and VHE gamma-ray emitters. Relativistic jets launched from the close environment of black holes in AGNs are known to produce powerful, variable HE and VHE gamma-ray emission in blazars \citep{cras-agn}. Furthermore, the detection in several microquasar jets of synchrotron emission extending from radio to X-ray energies unambiguously proves that TeV electrons are present on parsec scales, as in AGN jets \cite{2002Sci...298..196CBIS}.

There are currently two microquasars detected in HE gamma rays with both {\em AGILE} and the {\em Fermi}-LAT. \cyg\ \cite{2009Sci...326.1512F,2012arXiv1207.6288P,2012MNRAS.421.2947CBIS} and Cyg~X-1 \cite{2013arXiv1307.3264B,Sabatini:2013aa,2013arXiv1305.5920M} are well-known systems composed of a black hole candidate in orbit with a massive star. Their X-ray luminosity is around $10^{30}$ to $10^{31}$\,W, 3 to 4 orders-of-magnitude greater than in gamma-ray binaries with massive companions. This emission arises from the accretion disk and its corona, with a  contribution from the jet in some models \cite{Done:2007by}. The presence of relativistic jets in both systems is well-documented from radio observations, tracing non-thermal synchrotron emission from the ejecta. The activity of these jets is closely related to the X-ray spectral state \cite{Fender:2002hk}. The level of gamma-ray emission does not exceed $\approx 1$\% to 10\% of the X-ray luminosity. VHE observations have yielded only upper limits, except for a single flare event during MAGIC observations of Cyg X-1 with a significance of 4.9$\sigma$ (pre-trial) \cite{Albert:2007uw}. Subsequent long campaigns by the MAGIC and VERITAS collaborations did not detect another such VHE flare (hence the question mark in Table 1) \cite{2009arXiv0908.0714G,2011arXiv1110.1581Z}. \cyg\ has not been detected in VHE gamma rays \cite{2013ApJ...779..150A}.

The major feature of the HE gamma-ray emission is that it is related to the spectral state of the binary. Gamma rays are detected only when radio emission is present, clearly associating this high-energy emission to the relativistic jet. Spectral modeling also supports this view: the HE gamma-ray emission is distinct from the non-thermal power law spectral tails detected up to several MeV in both systems. The GeV emission is not a simple extrapolation of the MeV emission. The MeV tails probably arise from electrons in the corona while the HE gamma-ray emission arises further out \cite{2012MNRAS.426.1031Z,2014MNRAS.442.3243Z}. The orbital modulation of the HE emission from \cyg\ also constrains the location of the particles in the jet to be at a distance comparable to the orbital separation, very far from where most of the gravitational energy is released \cite{2010MNRAS.404L..55D,2012MNRAS.421.2956Z} ! This interpretation assumes that the modulation arises from inverse Compton scattering of photons from the massive star by electrons in the jet. HE gamma ray emission too close to the black hole would be strongly attenuated by pair production with the X-ray photons from the accretion disk \cite{1993A&A...278..307B,2011A&A...529A.120C}; HE gamma ray emission too far from the black hole would hardly be modulated and would require too high a power in electrons to generate the gamma-ray flux against the decreased stellar radiation density  \cite{2010MNRAS.404L..55D}. 

Particle acceleration at a recollimation shock within the jet has been proposed as an explanation for the distant gamma-ray emission site in \cyg. Such shocks occur when the pressure from the expanding jet has dropped to match the pressure of the surrounding medium. Recollimation shocks may account  for knots of emission in AGN jets \citep{cras-agn}, notably in M87 \cite{Stawarz:2006oh}, as particles in the jet are re-energized when they pass through such a shock. The Wolf-Rayet companion in \cyg\ has a mass loss rate and the tight orbit (4.8 hr) places the compact object only one stellar radius or so away from the star. The enormous ram pressure from the stellar wind is sufficient to cause the jet to recollimate within the binary system in \cyg, though not necessarily in Cyg X-1 (which has an O type companion). Attributing the gamma-ray emission to a specific location may not be necessary: in models where non-thermal electrons are continuously accelerated and cooled by inverse Compton scattering on stellar photons, the jet gamma-ray emission naturally peaks at a height comparable to the orbital separation \cite{2014MNRAS.440.2238Z}.

Despite extensive searches, HE or VHE gamma-ray emission has not been detected yet from other X-ray binaries \cite{2008ICRC....2..637D,2009arXiv0908.0714G,2011arXiv1110.1581Z,2015MNRAS.449.1686A,2015A&A...576A..36A}, including systems with low-mass companions \cite{2009A&A...508.1135H,2011ApJ...735L...5A,2014AA...568A.109A} and accreting neutron stars with high-mass companions \cite{2010ApJ...721..843A,2011arXiv1103.3250V}. In the latter, there is no radio emission from a relativistic jet and no pulsar wind. In the former, despite the presence of relativistic jets attested by radio observations, the radiation field from the low-mass star is much weaker. Particles in the jet may still be able to upscatter their synchrotron radiation to gamma-ray energies, but with a lower efficiency and closer to the base of the jet, where pair production may attenuate most of the emission. At very large distances, the dissipation of the jet kinetic energy in the interstellar medium though a shock may lead to detectable gamma-ray emission. An example is the recent report of HE gamma-ray emission from the microquasar SS 433 \cite{2015ApJ...807L...8B}. This detection, in a region of complex diffuse gamma-ray emission, remains to be confirmed.

Energy released by accretion also powers relativistic jets and gamma-ray emission in AGN and in gamma-ray bursts \citep{cras-agn}. Yet, it is only in microquasars that the release of energy in the different channels can be followed simultaneously, on reasonable timescales (changes occur on timescales of hours to days), albeit at the price of complex multiwavelength campaigns combining multiple facilities \cite{2012MNRAS.421.2947CBIS}: soft X-ray trace the accretion disk, hard X-rays trace its corona, radio waves trace the large-scale relativistic jet, and gamma rays trace the non-thermal particle injection during ejection (Figure 3). Although rare, gamma-ray emission from microquasars thus provides unique insights into the connection between accretion and ejection. 

\begin{figure}
\centering\resizebox{0.67\hsize}{!}{\includegraphics{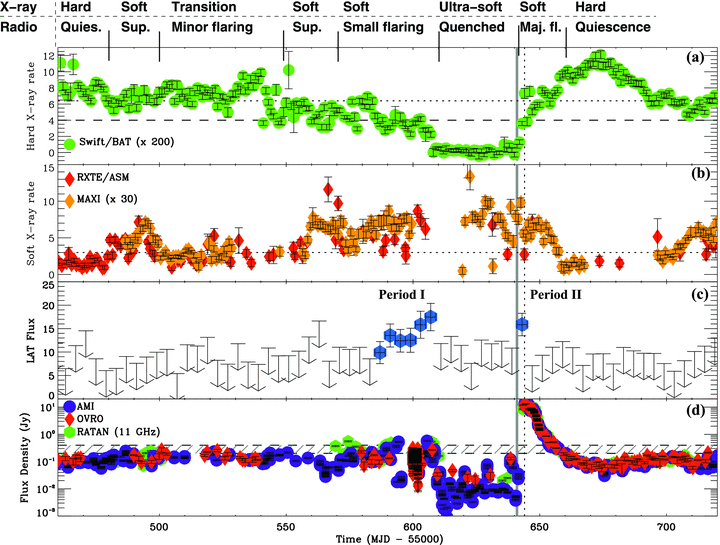}}
\caption{Multiwavelength lightcurve of the microquasar Cyg X-3  from 2010 September 21 to 2011 June 8, around the time of a giant radio flare (grey band shows the onset). The panels show (a) hard X-rays from {\em Swift}/BAT, (b) soft X-rays from {\em RXTE}/ASM and MAXI, (c) {\em Fermi}-LAT HE gamma rays, (d) 15 GHz radio flux (note the log scale). Roughly, the soft X-rays trace the accretion disk, hard X-rays trace the disk corona, gamma rays trace particle acceleration in the jet, and the radio trace the non-thermal emission from the relativistic jet. Figure reprinted from \citep{2012MNRAS.421.2947CBIS}, with permission of Oxford University Press, on behalf of the RAS.}
\end{figure}

\section{Novae and colliding wind binaries}
\label{bs}

\subsection{Eta Car and the missing gamma-ray emission from colliding wind binaries}

Colliding wind binaries (CWB) involve interacting winds, exactly like (high-mass) gamma-ray binaries (Fig. 1) except that the winds both originate from massive stars, hence the shock is non-relativistic \cite{2005xrrc.procE2.01P}. There is actually an evolutionary sequence from colliding wind binary, to gamma-ray binary (once one of the massive stars explodes and forms a neutron star), to X-ray binary (once the neutron star rotation has slowed down and the pulsar wind pressure is insufficient to prevent accretion). Non-thermal radio synchrotron emission has established that particle acceleration occurs in CWB  \cite{2013arXiv1308.3149D}. The large photon and matter densities in these systems  should easily lead to gamma-ray emission from inverse Compton scattering or pion-production \cite{2014ApJ...789...87R}. The gamma-ray emission would then inform on the maximum acceleration energy, the magnetic field, and the fraction of the wind kinetic energy converted to non-thermal energy at the shocks.

Despite expectations, gamma-ray emission is currently reported in only one CWB: Eta Car \cite{2009ApJ...698L.142T,2010ApJ...723..649A,2012MNRAS.424..128H,2012AA...544A..98R}. Eta Car is composed of two massive stars, one a gigantic $\approx 100\,\rm M_{\odot}$ luminous blue variable and the other an O or Wolf-Rayet type star, in a 5.5 year orbit \cite{2008MNRAS.384.1649D}. HE gamma rays are detected using {\em AGILE} and {\em Fermi}-LAT up to 100 GeV, though the system has not yet been detected at VHE by ground-based Cherenkov arrays, suggesting a spectral cutoff around a few 100 GeV \cite{2012MNRAS.424..128H}. The gamma-ray spectrum consists of two components: soft emission cutting off below 10 GeV and a hard power-law tail above 10 GeV, modulated on the orbital period \cite{2012AA...544A..98R}. 

The gamma-ray luminosity of Eta Car represents only $\approx 0.2\%$ of the kinetic power available from the stellar winds ($\approx 6\times 10^{30}\rm\,W$). Curiously, no gamma rays have been detected from other nearby CWB with comparable or even higher kinetic powers \cite{2013A&A...555A.102W}, so their gamma-ray efficiency must be lower. Perhaps this is due to circumstances in the binary environment that reduce the high-energy emission ({\em e.g.} strong radiative losses, or the shocked fraction of the wind is small). Perhaps this is signaling that diffusive acceleration is impeded in CWB compared to supernova remnants ({\em e.g.}  because the CWB shock structure is so unstable that the particles do not see the sharp interface required for Fermi acceleration).

\subsection{Novae}
In the initial stages of a nova eruption, a fraction of the white dwarf envelope is ejected while the rest, supported by the pressure from thermonuclear burning at the base, inflates to $\sim 100\rm\,R_{\odot}$ on a timescale of a day, followed by a phase of strong radiatively-driven mass loss lasting tens of days \cite{1994ApJ...437..802K}. Novae are powered by nuclear burning so low-energy ($\sim$1-10 MeV) gamma-ray emission from radioactive decay products has long been expected from these systems. However, gamma-ray emission above 100 MeV came as a surprise, because there was no compelling reason -- observational or theoretical --  to expect a lot of particle acceleration to occur (but see \cite{2007ApJ...663L.101T}).  

Five novae have now been detected in HE gamma rays with the {\em Fermi}-LAT \cite{2014Sci...345..554A}. The first, V407 Cyg, occurred in a binary composed of a red giant in a long orbit around the white dwarf (such systems are called ``symbiotic'' binaries) \cite{2010Sci...329..817A}. Red giants lose significant amounts of matter, $\sim 10^{-7}$ to $10^{-6}\rm\,M_{\odot}\,yr^{-1}$, in slow stellar winds with velocities of $\approx 20\rm\,km\,s^{-1}$. Thus, the nova ejecta had to propagate in a dense circumstellar environment,  forming an expanding shock on scales comparable to the orbital separation (10 AU), where particles can be accelerated to high energies. The system behaves as a small-scale analogue of a supernova remnant \cite{2013A&A...551A..37M}. Another symbiotic nova (not listed in Table 1), V745 Sco, yielded only a weak 2 to 3$\sigma$ signal in February 2014, probably because of its 7.8 kpc distance, to compare to 2.7 kpc in V407 Cyg. Many symbiotics are recurrent novae and we can look forward to {\em e.g.} the eruptions of RS Oph (4.2 kpc) in 2021 (H.E.S.S. observations were attempted a couple of months after the last eruption in 2006, probably too late) and of T Cr b in 2026 (0.9 kpc) \cite{2010ApJS..187..275S} ! 

After the detection of V407 Cyg, the presence of a red giant wind seemed to be the necessary condition for shock formation and gamma-ray emission in a nova. This turned out to be a red herring as the other four gamma-ray detections of novae all occurred in systems where the companion is a low-mass star with a puny stellar wind, yet the HE gamma-ray observations showed similar behaviour. The HE gamma-ray spectra are soft, peak below 1 GeV (VHE observations have given only upper limits \citep{2015arXiv150205853S}), compatible with either inverse Compton scattering or pion decay. With a luminosity $\approx 3\times 10^{28}\rm\,W$, about 0.1\% of the kinetic energy of the ejecta is channeled into gamma rays. Their orbital periods are not established yet  \cite{2012CBET.3136....1W,2013ApJ...768L..26P,v339orb} but are likely to be of order of a fraction of a day, implying binary separations of a few solar radii. The envelope in these novae expands well beyond the binary system into the interstellar medium. However, the HE gamma-ray emission lasts $\approx 2$ weeks, leaving little time to sweep up material in the tenuous interstellar medium compared to supernova remnants (or symbiotics). 

Various observational features suggest the coexistence in novae of outflow components with different velocities \cite{2014MNRAS.442..713M}. ``Internal'' shocks between these components may be strong enough to enable particle acceleration, and dense enough to reprocess the X-ray thermal emission generated by shock-heating into the optical band, as in type IIn supernovae \citep{2014MNRAS.442..713M,2015MNRAS.450.2739M}. Higher ejecta velocities are expected from heavier white dwarfs since the critical pressure is reached for smaller amounts of material deposited on the surface by accretion \cite{1994ApJ...437..802K}. The companion can also play a role in shaping the ejecta and the formation of shocks \cite{2014Natur.514..339CBIS}. Perhaps, all novae produce HE gamma-rays and only the closest ones are detected \cite{2014xru..confE.149M}. 

Gamma-ray observations have revealed a new facet of the nova phenomenon, one that may turn out to be very profitable to study since it gives a handle on diffusive shock acceleration in an environment that differs from the classic laboratory of supernova remnants \cite{cras-pwn}. Simultaneously fitting the gamma-ray emission with the thermal emission produced by shock-heated material, sensitive to the amount of swept-up material and to the shock velocity, can prove very constraining for the  fraction of the particles and energy injected in the acceleration process \citep{2013A&A...551A..37M,2015MNRAS.450.2739M}. This hope could be dashed if the origin of the shock(s) cannot be pinpointed precisely.

\section{What's next ?}
\label{conclusion}

The gamma-ray emission detected in binaries touches on many sub-fields of high-energy astrophysics. In terms of processes, it touches on the physics of magnetized, relativistic outflows in gamma-ray binaries and microquasars, where it informs and can be informed by our understanding of gamma-ray bursts and active galactic nuclei ; it touches on diffusive shock acceleration in colliding wind binaries and novae, where it is comparable to observations and theories of supernova remnants or cluster shocks. Gamma-ray observations of binaries may yield irreplaceable insights: for instance, if inverse Compton gamma rays from upscattered stellar photons do trace the free pulsar wind (unobservable in PWN), or when gamma rays sign non-thermal processes in CWB (if deeply-embedded radio synchrotron emission is hidden by free-free absorption in the stellar winds) or novae (where high ejecta densities may hide the thermal emission from the shock-heated material \citep{2015MNRAS.450.2739M}). 

The variety of binaries emitting gamma rays proves that particle acceleration is much more widespread than would have been considered a decade ago. At present, this variety is limited to the HE gamma ray regime but there is no reason to expect less  in the VHE regime. With a ten-fold increase in sensitivity, the Cherenkov Telescope Array (CTA, \citep{cras-future}) will enable us to see deeper below the tip of the iceberg (detecting, for instance, VHE emission from low-mass gamma-ray binaries,  flaring microquasars, or CWB \cite{2012arXiv1210.3215P}). VHE information is essential to constrain the maximal particle energy and characterize the efficiency of the accelerator \cite{Khangulyan:2007me}. For all we know, binaries may even add the odd contribution to the sea of Galactic cosmic rays \cite{Heinz:2002qb}. One may also hope that these sensitive VHE observations will uncover emission from unexpected sources, including binaries. Faint, hard gamma-ray sources could be lost in the Galactic diffuse emission in the {\em Fermi}-LAT band and pop up in a Galactic plane survey with CTA.  For instance, \hessj\ has not been detected in HE gamma rays despite being a bright VHE source \cite{2013MNRAS.436..740CBIS}. One lesson to be learned from the present observations is that the HE and VHE regimes are rarely simple extrapolations of each other.

Variability is a main feature of gamma-ray emission from binaries. All of the sources in Table 1 are variable. Characterizing this variability constrains the timescales of the acceleration and radiative processes that are involved. Significant progress can be expected with CTA \citep{cras-future}. Its sensitivity accross a broad band of energies, from a few tens of GeV to tens of TeV, will enable high quality spectral energy distributions to be reconstructed on much shorter timescales than possible now, allowing {\em e.g.} for exquisite phase-resolved studies of the emission from gamma-ray binaries. However, compared to other objects, there is no escape from the additional efforts and perseverance required to obtain a complete observational picture, identify the different regimes of gamma-ray emission, accumulate high signal-to-noise spectra for each regime, establish the relationships to variability at other wavelengths, and measure the basic binary parameters when the system is poorly known (nature of the components, orbital period...). 

Finally, on the longer term, the frontier of gamma-ray observations will have to move to lower energies, in the 1-100 MeV range (LE gamma rays) \citep{cras-future}. 
This range is undoubtedly rich in sources associated with binaries. Emission can be expected from the very hot plasma in accretion disk coronae, from non-thermal particle injection, from radioactive decay or spallation. 
The current sensitivity in LE gamma rays is poor compared to the HE/VHE gamma-ray regimes and compared to the neighbouring hard X-ray regime, where the unprecedent sensitivity of {\em NuSTAR} observations is providing a wealth of discoveries and fresh insights \citep{2013ApJ...770..103H}. 
In Table 1, \ls, \lsi, \cyg, Cyg X-1 are already known emitters of LE gamma rays but the quality of the spectra and lightcurves is far from those at other wavelengths. 
A ten-fold improvement in sensitivity in the LE range would certainly revolutionize our understanding of how particles in various objects, including binaries, cross from the thermal to the non-thermal pool.

%{\noindent \bf Acknowledgements.} 
\section*{Acknowledgements}
I thank the Centre National d'Etudes Spatiales (CNES) for continued support.
%% References


%\bibliographystyle{elsarticle-num}
%\bibliographystyle{aa}       % A&A
\bibliographystyle{unsrtnat}
\bibliography{../../../BIBLIO.bib}


\end{document}